\documentclass[aps,prd,onecolumn,groupedaddress,showpacs,nofootinbib,amssymb]{revtex4}
\usepackage{amsmath}
\usepackage{amssymb}
\usepackage{amsfonts}
\usepackage{graphicx,bm}
\usepackage{dcolumn}
\usepackage{epsfig}
\usepackage{color,amsxtra}
\usepackage{epsf}
\usepackage{enumerate}
\usepackage{hhline}
\usepackage{array}
\usepackage{tabularx}
\newcommand{\be}{\begin{equation}}
\newcommand{\ee}{\end{equation}}
\newcommand{\bea}{\begin{eqnarray}}
\newcommand{\eea}{\end{eqnarray}}
\newcommand{\beaa}{\begin{eqnarray*}}
\newcommand{\eeaa}{\end{eqnarray*}}

\newcommand{\e}{\mathrm{e}}

\allowdisplaybreaks[4]

\newcommand{\Eqn}[1]{&\hspace{-0.2em}#1\hspace{-0.2em}&}

\def\be{\begin{equation}}
\def\ee{\end{equation}}
\def\bea{\begin{eqnarray}}
\def\eea{\end{eqnarray}}

\def\e{\mathrm{e}}

\begin{document}


\title{Bouncing cosmology in modified Gauss-Bonnet gravity}
\author{Kazuharu Bamba$^{1, 2}$,
Andrey N. Makarenko$^{3}$,
Alexandr N. Myagky$^{4}$ 
and 
Sergei D. Odintsov$^{3, 5, 6}$
}
\affiliation{
$^1$Leading Graduate School Promotion Center,
Ochanomizu University, 2-1-1 Ohtsuka, Bunkyo-ku, Tokyo 112-8610, Japan\\
$^2$Kobayashi-Maskawa Institute for the Origin of Particles and the
Universe, Nagoya University, Nagoya 464-8602, Japan\\
$^3$Tomsk State Pedagogical University, ul. Kievskaya, 60, 634061 Tomsk and \\
National Research Tomsk State University,  Lenin Avenue, 36, 634050 Tomsk, Russia\\
$^4$National Research Tomsk Polytechnic University, Lenin Avenue, 30,
634050, Tomsk, Russia\\
$^5$Instituci\`{o} Catalana de Recerca i Estudis Avan\c{c}ats (ICREA),
Barcelona, Spain\\
$^6$Institut de Ciencies de l'Espai (CSIC-IEEC),
Campus UAB, Facultat de Ciencies, Torre C5-Par-2a pl, E-08193 Bellaterra
(Barcelona), Spain
}


\begin{abstract}
We explore bounce cosmology in $F(\mathcal{G})$ gravity with the Gauss-Bonnet invariant $\mathcal{G}$. We reconstruct $F(\mathcal{G})$ gravity theory to
realize the bouncing behavior in the early universe and
examine the stability conditions for its cosmological solutions.
It is demonstrated that the bouncing behavior with an exponential as well 
as a power-law scale factor naturally occurs in modified Gauss-Bonnet 
gravity. We also derive the $F(\mathcal{G})$ gravity model to produce 
the ekpyrotic scenario. 
Furthermore, we construct the bounce with the scale factor composed of a 
sum of two exponential functions and show that not only the early-time 
bounce but also the late-time cosmic acceleration can occur in the 
corresponding modified Gauss-Bonnet gravity. 
Also, the bounce and late-time solutions in this unified model 
is explicitly analyzed. 
\end{abstract}

\hspace{13.1cm} OCHA-PP-318

\maketitle

\textit{Introduction}--
%
As a cosmological model to describe the early universe,
the matter bounce scenario~\cite{B-R-BCS} is known.
In this scenario, in the contraction phase the universe
is dominated by matter,
and a non-singular bounce occurs. Also, the density perturbations
whose spectrum is consistent with the observations
can be produced (for a review, see~\cite{Novello:2008ra}).
In addition, after the contracting phase,
the so-called BKL instability~\cite{Belinsky:1970ew}
happens, so that the universe will be anisotropic.
The way of avoiding this instability~\cite{Erickson:2003zm} and
issues of the bounce~\cite{X-S} in the Ekpyrotic scenario~\cite{Khoury:2001wf} has been investigated~\cite{Cai:2012va, Cai:2013vm}.
Moreover, the density perturbations
in the matter bounce scenario with two scalar fields
has recently been examined~\cite{Cai:2013kja}.

On the other hand, various cosmological observations support
the current cosmic accelerated expansion.
To explain this phenomenon in the homogeneous and isotropic universe,
it is necessary to assume the existence of dark energy, which
has negative pressure, or propose that gravity is modified on large scales
(for recent reviews on issues of dark energy and modified gravity theories,
see, e.g.,~\cite{R-NO-CF-CD, Bamba:2012cp}).
Regarding the latter approach, there have been proposed a number of
modified gravity theories such as $F(R)$ gravity.
The bouncing behavior has been investigated
in $F(R)$ gravity~\cite{OS-SZ, BouhmadiLopez:2012qp, 
Leon:2013bra,Bamba:2013fha},
string-inspired gravitational theories~\cite{Biswas:2005qr},
non-local gravity~\cite{BKMV-BKM}. 
A relation between the bouncing behavior and 
the anomalies on the cosmic microwave background (CMB) radiationhas also been discussed~\cite{Liu:2013kea}.

In this Letter, we explore bounce cosmology in $F(\mathcal{G})$ gravity with
$F(\mathcal{G})$ an arbitrary function of the Gauss-Bonnet invariant
$\mathcal{G}=R^{2}-4R_{\mu\nu}R^{\mu\nu}+
R_{\mu\nu\rho\sigma}R^{\mu\nu\rho\sigma}$, where
$R_{\mu\nu}$ is the Ricci tensor and $R_{\mu\nu\rho\sigma}$ is
the Riemann tensor.
Such $F(\mathcal{G})$ theory has been proposed as 
gravitational alternative for dark energy and 
inflation in Ref.~\cite{Nojiri:2005jg} 
and its application to the late-time cosmology~\cite{Cognola:2006eg}
has been studied.
Moreover, cosmology in a theory with a dynamical dilaton coupling to
the Gauss-Bonnet invariant has also been studied~\cite{NOS-BGO}.
We use units of $k_\mathrm{B} = c_{\mathrm{l}} = \hbar = 1$,
where $c$ is the speed of light, and denote the
gravitational constant $8 \pi G$ by
${\kappa}^2 \equiv 8\pi/{M_{\mathrm{Pl}}}^2$
with the Planck mass of $M_{\mathrm{Pl}} = G^{-1/2} = 1.2 \times
10^{19}$\,\,GeV.

In the following, we first explain $F(\mathcal{G})$ gravity
and its reconstruction method.
We also investigate the stability of the solutions
in the reconstructed $F(\mathcal{G})$ gravity model.
As more concrete examples,
we study an exponential model and a power-law model,
in which the bouncing behavior happens. 
In addition, the $F(\mathcal{G})$ gravity model to make 
the ekpyrotic scenario~\cite{Enc} is build. 
Next, we examine a sum of two exponential functions model
of the scale factor.
We explicitly show that in this model
the bouncing behavior in the early universe
and the late-time cosmic acceleration can be
realized in a unified way. 
Furthermore, we make the stability analysis of 
the bounce and late-time solutions in the unified model. 
Finally, our results are summarized.

\textit{$F(\mathcal{G})$ theory of gravity}--
%
The action of $F(\mathcal{G})$ gravity model is described as 
\cite{Nojiri:2005jg}
\begin{equation}
S = \frac{1}{2\kappa^2} \int d^4 x \sqrt{-g} \left(R+F(\mathcal{G})\right)
+S_{\mathrm{matter}}\,,
\label{eq:1}
\end{equation}
where $g$ is the determinant of the metric tensor $g_{\mu\nu}$
and $S_{\mathrm{matter}}$ is the matter action.
It follows from this action that
the gravitational field equation reads
\begin{eqnarray}
&&
R_{\mu\nu}-\frac{1}{2} g_{\mu\nu} R - \frac{1}{2} g_{\mu\nu}
F(\mathcal{G})
-\left(
-2RR_{\mu\nu} +4R_{\mu\rho}R_{\nu}{}^{\rho}
-2R_{\mu}{}^{\rho\sigma\tau}R_{\nu\rho\sigma\tau}
+4g^{\alpha\rho}g^{\beta\sigma}R_{\mu\alpha\nu\beta}R_{\rho\sigma}
\right) F'(\mathcal{G})
\nonumber \\
&&
{}
-2\left({\nabla}_{\mu}{\nabla}_{\nu} F'(\mathcal{G}) \right)R
+2g_{\mu \nu}\left(\Box F'(\mathcal{G})\right)R
-4\left(\Box F'(\mathcal{G}) \right)R_{\mu \nu}
+4\left({\nabla}_{\rho}{\nabla}_{\mu} F'(\mathcal{G}) \right)
R_{\nu}{}^{\rho}
+4\left({\nabla}_{\rho}{\nabla}_{\nu} F'(\mathcal{G}) \right)
R_{\mu}{}^{\rho}
\nonumber \\
&&
{}
-4g_{\mu \nu}\left({\nabla}_{\rho}{\nabla}_{\sigma}
F'(\mathcal{G}) \right)R^{\rho\sigma}
+4\left({\nabla}_{\rho}{\nabla}_{\sigma} F'(\mathcal{G}) \right)
g^{\alpha\rho}g^{\beta\sigma}R_{\mu\alpha\nu\beta}
= \kappa^2 T^{(\mathrm{matter})}_{\mu \nu}\,.
\label{eq:2}
\end{eqnarray}
Here, the prime denotes the derivative with respect to $\mathcal{G}$,
${\nabla}_{\mu}$ is the covariant derivative,
$\Box \equiv g^{\mu \nu} {\nabla}_{\mu} {\nabla}_{\nu}$
is the covariant d'Alembertian, and
$T^{\mu (\mathrm{matter})}_{\,\,\,\,\nu} = \mathrm{diag} \left(-\rho_{\mathrm{matter}}, p_{\mathrm{matter}}, p_{\mathrm{matter}}, p_{\mathrm{matter}}
\right)$
is the energy-momentum tensor of matter, where
$\rho_{\mathrm{matter}}$ and $p_{\mathrm{matter}}$ are the energy density and pressure of matter, respectively.
We take the flat Friedmann-Lema\^{i}tre-Robertson-Walker (FLRW) metric,
given by
\begin{equation}
ds^2 = - dt^2 + a^2(t) \sum_{i=1,2,3}\left(dx^i\right)^2\,,
\label{eq:3}
\end{equation}
where $a$ is the scale factor,
$H=\dot{a}/a$ is the Hubble parameter, and
the dot shows the time derivative.
In this background, we have $R = 6\dot{H} + 12H^2$
and $\mathcal{G} = 24H^2 \left( \dot{H} + H^2 \right)$.
The gravitational field equations become~\cite{Cognola:2007vq}
\begin{eqnarray}
&&
6H^2 + F(\mathcal{G}) - \mathcal{G} F'(\mathcal{G})
+ 24H^3 \dot{\mathcal{G}} F''(\mathcal{G})
= 2\kappa^2 \rho_{\mathrm{matter}}\,,
\label{eq:4} \\
&&
4\dot{H}+6H^2
+ F(\mathcal{G}) -\mathcal{G}F'(\mathcal{G}) +16H \dot{\mathcal{G}} \left( \dot{H} + H^2 \right) F''(\mathcal{G}) + 8H^2 \ddot{\mathcal{G}} F''(\mathcal{G}) + 8H^2 \dot{\mathcal{G}}^2 F'''(\mathcal{G})
= -2\kappa^2 p_{\mathrm{matter}}\,.
\label{eq:5}
\end{eqnarray}
In what follows, we investigate only gravity part of the action
in Eq.~(\ref{eq:1}) without its matter part.
For the case that the scale factor $a(t)$ has the form of linear combination
of two exponential terms as
\begin{equation}
a(t) = \sigma \exp(\lambda t) + \tau \exp(-\lambda t)\,,
\label{eq:6}
\end{equation}
where $\lambda (\neq 0)$, $\sigma$, and $\tau$ are constants.
In this case, we find
\begin{equation}
H(t) = \frac{\dot{a}}{a} = \lambda \frac{\sigma \exp(\lambda t) - \tau \exp(-\lambda t)}{\sigma \exp(\lambda t) + \tau \exp(-\lambda t)}\,,
\quad
\mathcal{G}(t) =
24 \lambda^4 \frac{\left( \exp( 2 \lambda t) \sigma -\tau \right)^2}{\left( \exp( 2 \lambda t) \sigma +\tau \right)^2}\,.
\label{eq:7}
\end{equation}
%

\textit{Reconstruction method of $F(\mathcal{G})$ gravity}--
%
Next, we reconstruct $F(\mathcal{G})$ gravity models
by using the method~\cite{R-M, Bamba:2008ut}.
Introducing proper functions $P(t)$ and $Q(t)$ of a scalar field $t$,
which is interpreted as the cosmic time,
the action in Eq.~({\ref{eq:1}}) without matter is described as
\begin{equation}
S=\frac{1}{2\kappa^2} \int d^{4}x \sqrt{-g}
\left( R+P(t)\mathcal{G}+Q(t) \right)\,.
\label{eq:8}
\end{equation}
By varying this action with respect to $t$, we obtain
$\left( d P(t)/dt \right) \mathcal{G}+\left(d Q(t)/dt\right)=0$.
Solving this equation in terms of $t$, we get $t=t(\mathcal{G})$.
The substitution of $t=t(\mathcal{G})$ into Eq.~(\ref{eq:8}) yields
$F(\mathcal{G})=P(t)\mathcal{G}+Q(t)$.
Using this equation and Eq.~(\ref{eq:4}), we find
\begin{equation}
Q(t) = -6H^2(t) -24 H^3(t) \frac{dP(t)}{dt}\,.
\label{eq:9}
\end{equation}
With this equation and the relation $F(\mathcal{G})=P(t)\mathcal{G}+Q(t)$,
we acquire
\begin{equation}
2H^2(t) \frac{d^2 P(t)}{d t^2} + 2H(t) \left( 2\dot{H}(t) -H^2(t) \right)
\frac{d P(t)}{d t} + \dot{H}(t) = 0\,.
\label{eq:10}
\end{equation}

Suppose the scale factor is given by Eq.~(\ref{eq:6}),
we present the general solution of Eq.~(\ref{eq:10})
for the following two cases.

\noindent
In {\bf Case 1}: $\lambda > 0$, $\sigma > 0$, $\tau > 0$.

The general solution of Eq.~(\ref{eq:10}) becomes
\begin{eqnarray}
P(t) \Eqn{=} c_1 +
\left( c_2 -\frac{1}{4\lambda^2 \sqrt{\sigma\tau}} \arctan \left( \e^{\lambda t} \sqrt{\frac{\sigma}{\tau}}\right)
\right) \mathcal{I} -\frac{1}{2\lambda^2} \ln \left( 1-\frac{2\tau}{\e^{2\lambda t} \sigma +\tau} \right)\,,
\label{eq:11} \\
\mathcal{I} \Eqn{\equiv}
\frac{\e^{3\lambda t} \sigma^2 -6\e^{\lambda t}\sigma\tau +\e^{-\lambda t}\tau^2}{\e^{2\lambda t} \sigma -\tau}\,,
\label{eq:12}
\end{eqnarray}
where $c_1$ and $c_2$ are constants. {}From Eq.~(\ref{eq:9}), we obtain
\begin{equation}
Q(t) = -6\lambda^2 \e^{-\lambda t} \left( \e^{2\lambda t}\sigma -\tau \right)
\left( 4c_2 \lambda^2 - \frac{1}{\sqrt{\sigma\tau}}
\arctan \left( \e^{\lambda t} \sqrt{\frac{\sigma}{\tau}}\right)
\right)\,.
\label{eq:13}
\end{equation}
Plugging this expression with Eq.~(\ref{eq:9}), we have
\begin{equation}
t_{\pm} \equiv \frac{1}{2\lambda} \ln \left( - \frac{\left( \mathcal{G} \pm 4\sqrt{6} \sqrt{\mathcal{G}} \lambda^2 + 24 \lambda^4 \right) \tau}{\left( \mathcal{G} -24\lambda^4 \right)\sigma} \right)\,,
\quad
0 \leq \mathcal{G} < 24\lambda^4 \,.
\label{eq:14}
\end{equation}
Accordingly, by solving $F(\mathcal{G})=P(t)\mathcal{G}+Q(t)$, we find
the most general form of $F(\mathcal{G})$ as
\begin{equation}
F_\pm^{(1)}(\mathcal{G}) = c_1 \mathcal{G}
+ c_2 \sqrt{\mathcal{G} \left( 24 \lambda^4 - \mathcal{G} \right)}
-\frac{1}{2\lambda^2} \mathcal{G}
\ln \left( \pm \frac{\sqrt{\mathcal{G}}}{2\sqrt{6} \lambda^2} \right)
\pm \frac{1}{\lambda^2} \sqrt{\mathcal{G} \left( 24 \lambda^4 - \mathcal{G} \right)} \arctan \left( \frac{\pm\sqrt{\mathcal{G}} +2\sqrt{6} \lambda^2}{\sqrt{24 \lambda^4 - \mathcal{G}}} \right)\,,
\label{eq:15}
\end{equation}
%
where the superscription $(1)$ of $F_\pm^{(1)}(\mathcal{G})$ means ``Case $1$''.
Note that the functions $F_{+}^{(1)}(\mathcal{G})$ and
$F_{-}^{(1)}(\mathcal{G})$ are defined for $0\leq \mathcal{G}<24\lambda^4$
and the function $F_{-}^{(1)}(\mathcal{G})$ takes only complex values.
Furthermore, we see that
\begin{equation}
\lim_{\mathcal{G} \to +0} F_+^{(1)} (\mathcal{G}) = 0\,,
\quad
\lim_{\mathcal{G} \to 24 \lambda^4 -0} F_+^{(1)} (\mathcal{G})
= 24 c_1 \lambda^4\,.
\label{eq:16}
\end{equation}

\noindent
In {\bf Case 2}: $\lambda > 0$, $\sigma > 0$, $\tau < 0$.

The general solution of Eq.~(\ref{eq:10}) is derived as
\begin{equation}
P(t) = c_1 + c_2 \mathcal{I} +
\frac{\e^{-\lambda t}}{4\lambda^2 \left(\e^{2\lambda t} \sigma - \tau \right) \sqrt{-\sigma\tau}} \left(\e^{4\lambda t} \sigma^2 -6\e^{2\lambda t} \sigma\tau +\tau^2 \right) \arctan \left( \e^{\lambda t} \sqrt{-\frac{\sigma}{\tau}}\right)
+ \frac{1}{\lambda^2} \arctan \left( \e^{-2\lambda t} \frac{\tau}{\sigma}
\right)\,.
\label{eq:17}
\end{equation}
Using Eq.~(\ref{eq:9}), we get
\begin{equation}
Q(t) = -6\lambda^2 \e^{-\lambda t} \left( \e^{2\lambda t}\sigma -\tau \right)
\left( 4c_2 \lambda^2 + \frac{1}{\sqrt{-\sigma\tau}}
\arctan \left( \e^{\lambda t} \sqrt{-\frac{\sigma}{\tau}}\right)
\right)\,.
\label{eq:18}
\end{equation}
With the equation
$\left( d P(t)/dt \right) \mathcal{G}+\left(d Q(t)/dt\right)=0$,
we observe that for $\mathcal{G} > 24 \lambda^4$,
the expression of $t_\pm$ is given by Eq.~(\ref{eq:14}).
It follows from solving the equation
$F(\mathcal{G})=P(t)\mathcal{G}+Q(t)$ that
the most general form of $F(\mathcal{G})$ reads
\begin{equation}
F_\pm^{(2)}(\mathcal{G}) = c_1 \mathcal{G}
+ c_2 \sqrt{\mathcal{G} \left( 24 \lambda^4 - \mathcal{G} \right)}
-\frac{1}{\lambda^2} \mathcal{G}
\arctan \left( \frac{\sqrt{\mathcal{G}} \mp 2\sqrt{6} \lambda^2}{\sqrt{\mathcal{G}} \pm 2\sqrt{6} \lambda^2} \right)
+ \frac{1}{\lambda^2} \sqrt{\mathcal{G} \left( \mathcal{G}-24 \lambda^4  \right)} \arctan \left( \frac{\sqrt{\mathcal{G}} \pm2\sqrt{6} \lambda^2}{\sqrt{\mathcal{G} -24 \lambda^4}} \right)\,.
\label{eq:19}
\end{equation}
%
However, it is clear that the function $F_\pm^{(2)} (\mathcal{G})$ has no real values for $\mathcal{G} > 24 \lambda^4$.

\textit{Stability of the solutions}--
%
We define $H^2(t) \equiv \tilde{g}(N)$, where the number of $e$-fold is
defined by $N \equiv \ln \left(a/a_{\star} \right) (\geq 0)$ with $a_{\star} = a(t_{\star})$ the scale factor at a fiducial time $t_{\star}$. Using $\tilde{g} (N)$, the Friedmann equation (\ref{eq:4}) is
rewritten to
\begin{equation}
6\tilde{g}(N) + F(\mathcal{G}) -12\tilde{g}(N) \left( \tilde{g}'(N) + 2\tilde{g}(N) \right) F'(\mathcal{G}) +288\tilde{g}^2(N) \left[
\left(\tilde{g}'(N)\right)^2 + \tilde{g}(N) \tilde{g}''(N) +4\tilde{g}(N) \tilde{g}'(N) \right] F''(\mathcal{G})
= 0\,.
\label{eq:20}
\end{equation}
Here, the prime shows the derivative with respect to $N$,
we have used $\mathcal{G} = 12 \tilde{g}(N) \left( \tilde{g}'(N) + 2\tilde{g}(N)\right)$, and only the gravity part has been considered
(i.e., $\rho_{\mathrm{matter}} = 0$).
We express the solution of Eq.~(\ref{eq:4}) as
$\tilde{g} = \tilde{g}_0(N)$. To examine the stability of this background solution, we describe $\tilde{g}(N) = \tilde{g}_0(N) + \delta \tilde{g}(N)$, where $\delta \tilde{g}(N)$ is the perturbation around the background solution.
By substituting the above expression of $\tilde{g}(N)$ into Eq.~(\ref{eq:20}),
we obtain
\begin{equation}
\mathcal{J}_1 \delta \tilde{g}''(N)
+ \mathcal{J}_2 \delta \tilde{g}'(N) + \mathcal{J}_3 \delta \tilde{g}(N) =0\,,
\label{eq:21}
\end{equation}
with
\begin{eqnarray}
\mathcal{J}_1 \Eqn{\equiv} 288 \tilde{g}_0^3(N) F''(\mathcal{G}_0)\,,
\label{eq:22} \\
\mathcal{J}_2 \Eqn{\equiv} 432 \tilde{g}_0^2(N)
\left\{ \left( 2 \tilde{g}_0 (N) + \tilde{g}_0' (N) \right) F''(\mathcal{G}_0) +8\tilde{g}_0 (N) \left[
\left( \tilde{g}_0' (N) \right)^2 + \tilde{g}_0 (N) \left( 4\tilde{g}_0' (N) +
\tilde{g}_0'' (N) \right) \right] F'''(\mathcal{G}_0) \right\}\,,
\label{eq:23} \\
\mathcal{J}_3 \Eqn{\equiv}
6 \left( 1 +24\tilde{g}_0 (N)
\left\{
\left[ -8 \tilde{g}_0^2 (N) +3\left( \tilde{g}_0' (N) \right)^2
+6\tilde{g}_0 (N) \left( 3\tilde{g}_0' (N) + \tilde{g}_0'' (N) \right)
\right] F''(\mathcal{G}_0)
\right. 
\right.
\nonumber \\
&&
{} +\left. 
\left.
24\tilde{g}_0 (N) \left( 4\tilde{g}_0 (N) + \tilde{g}_0' (N) \right)
\left[ \left( \tilde{g}_0' (N) \right)^2  + \tilde{g}_0 (N)
\left( 4\tilde{g}_0' (N) +\tilde{g}_0'' (N) \right)
\right] F'''(\mathcal{G}_0)
\right\}
\right)\,,
\label{eq:24}
\end{eqnarray}
where $\mathcal{G}_0 \equiv 12 \tilde{g}_0 (N) \left( \tilde{g}_0' (N) +2 \tilde{g}_0 (N) \right)$.
Therefore, the stability conditions are given by
$\mathcal{J}_2/\mathcal{J}_1 >0$ and $\mathcal{J}_3/\mathcal{J}_1 >0$.

We investigate a model with $F(\mathcal{G}) = F_+^{(1)} (\mathcal{G})$
in Eq.~(\ref{eq:15}).
As an example, we explore a bouncing solution with the scale factor
\begin{equation}
a(t) = \cosh \left( \lambda t \right) =
\frac{\exp \left( \lambda t \right) + \exp \left( -\lambda t \right)}{2}\,.
\label{eq:25}
\end{equation}
In this model, we have
\begin{equation}
N = \ln \cosh \left( \lambda t \right)\,,
\quad
H = \dot{N} = \lambda \tanh \left( \lambda t \right)\,,
\label{eq:26}
\end{equation}
where we have taken $a_{\star} = 1$. {}From these expressions, we find
\begin{equation}
\tilde{g}_0 (N) = H^2 (N) = \lambda^2 \left(1-\exp \left( -2N \right)
\right)\,,
\quad
\mathcal{G}_0 = 12 \tilde{g}_0 (N) \left( \tilde{g}_0' (N) +2 \tilde{g}_0 (N) \right) = 48 \lambda^4 \exp \left( -N \right) \sinh \left( N \right)\,.
\label{eq:27}
\end{equation}
Thus, the stability conditions for this model with the scale factor in
  Eq.~(\ref{eq:25}) reads
\begin{eqnarray}
&&
\frac{\mathcal{J}_2}{\mathcal{J}_1} =
\frac{3}{2}\left( 1 + \coth (N) \right) +48 \lambda^4 \exp \left( -2N \right)
\frac{F'''(\mathcal{G}_0)}{F''(\mathcal{G}_0)} >0\,.
\label{eq:28} \\
&&
\frac{\mathcal{J}_3}{\mathcal{J}_1} =
\frac{4-18\e^{2N}+18\e^{4N}-4\e^{6N}}{\left(-1+\e^{2N} \right)^3} + \frac{\e^{6N}}{48 \lambda^6 \left(-1+\e^{2N} \right)^3} \frac{1}{F''(\mathcal{G}_0)}
+\frac{96 \lambda^4 \e^{-2N} \left( -1+2\e^{2N} \right)}{-1+\e^{2N}} \frac{F'''(\mathcal{G}_0)}{F''(\mathcal{G}_0)} >0\,.
\label{eq:29}
\end{eqnarray}
Consequently, we observe that
\begin{equation}
\lim_{N \to +\infty} \frac{\mathcal{J}_2}{\mathcal{J}_1} = 6 \,,
\quad
\lim_{N \to +\infty} \frac{\mathcal{J}_3}{\mathcal{J}_1} = 8 \,.
\label{eq:30}
\end{equation}
In most cases, it is easy to find $N_0 = N_0 (c_2, \lambda)$ and for all $N > N_0$ both stability conditions will be executed.
Thus, for the model of $F(\mathcal{G}) = F_+^{(1)} (\mathcal{G})$
in Eq.~(\ref{eq:15}) with the scale factor in Eq.~(\ref{eq:25}),
the background solution is stable.

Here, we mention finite-time future singularities~\cite{Nojiri:2005sx, Bamba:2008ut} in $F(\mathcal{G})$ gravity by following the observations
in Ref.~\cite{Bamba:2009uf}.
In the limit of a finite time $t_{\mathrm{s}} (=\mathrm{constant} > t)$
in the future, cosmological quantities of the scale factor $a(t)$, the effective (namely, total) energy density $\rho_{\mathrm{eff}}$ and pressure $P_{\mathrm{eff}}$ of the universe, and the higher derivative of the Hubble parameter would diverge.
The finite-time future singularities can be classified into four types~\cite{Nojiri:2005sx}.
In the limit of $t\to t_{\mathrm{s}}$,
(a) Type I (``Big Rip''):\
$a \to \infty$,
$\rho_{\mathrm{eff}} \to \infty$, and
$| P_{\mathrm{eff}} | \to \infty$.
This type envolves the case that
$\rho_\mathrm{{eff}}$ and $P_{\mathrm{eff}}$ at $t_{\mathrm{s}}$ are finite.
(b) Type II (``sudden''):\
$a \to a_{\mathrm{s}}$,
$\rho_{\mathrm{eff}} \to \rho_{\mathrm{s}}$, and
$| P_{\mathrm{eff}} | \to \infty$,
where $a_{\mathrm{s}} (\neq 0)$ and $\rho_{\mathrm{s}}$ are constants.
(c) Type III:\
$a \to a_{\mathrm{s}}$,
$\rho_{\mathrm{eff}} \to \infty$, and
$| P_{\mathrm{eff}} | \to \infty$.
(d) Type IV:\
$a \to a_{\mathrm{s}}$,
$\rho_{\mathrm{eff}} \to 0$, $| P_{\mathrm{eff}} | \to 0$,
but higher derivatives of $H$ diverge.
This type also includes
the case that $\rho_{\mathrm{eff}}$ and/or $| P_{\mathrm{eff}} |$
are finite at $t = t_{\mathrm{s}}$.

If $H = \bar{h}/\left( t_{\mathrm{s}} -t \right)$ with
$\bar{h} (>0)$ a positive constant,
the reconstructed form of $F(\mathcal{G})$ to produce
the Big Rip singularity is given by $F(\mathcal{G}) =
\left[\sqrt{6\bar{h}\left(1+\bar{h}\right)}/{\left(1-\bar{h}\right)}\right]
\sqrt{\mathcal{G}}+d_{1}\mathcal{G}^{\left(h+1\right)/4}+d_{2}\mathcal{G}$,
where $d_1$ and $d_2$ are constants.
When $\bar{h}=1$, we find $F(\mathcal{G}) = \left(\sqrt{3}/2 \right)\sqrt{\mathcal{G}} \ln \left(\zeta \mathcal{G} \right)$, where $\zeta (>0)$ is a positive constant.
In the case of large values of $\mathcal{G}$,
$F(\mathcal{G}) \sim \xi \sqrt{\mathcal{G}} \ln(\zeta \mathcal{G})$
with $\xi (>0)$ a positive constant,
and eventually the Big Rip singularity happens.
The same consequence is obtained for
$F(\mathcal{G}) \sim \xi \sqrt{\mathcal{G}} \ln(\zeta \mathcal{G}^{u}+
\mathcal{G}_\mathrm{c})$, where
$u (>0)$ and $\mathcal{G}_\mathrm{c}$ are constants.

Moreover, when $H = \bar{h}/\left( t_{\mathrm{s}} -t \right)^{\upsilon}$ with $\upsilon$ a constant, for $\upsilon >1$,
if the value of $\mathcal{G}$ is large,
$F(\mathcal{G}) \sim -l_2 \sqrt{\mathcal{G}}$ with $l_2 >0$,
there occurs the Type I singularity.
For $0 <\upsilon <1$,
when $\mathcal{G}$ is large, we acquire
$F(\mathcal{G}) \sim l_3 |\mathcal{G}|^{v}$ with
$l_3 (>0)$ a positive constant and
$v = 2\upsilon/\left(3\upsilon+1\right)$, where $0< v <1/2$.
Accordingly, the Type III singularity appears.
When $\mathcal{G} \to - \infty$,
$F(\mathcal{G}) \sim l_3 |\mathcal{G}|^{v}$, where
$-\infty<v<0$, and therefore $-1/3 < \upsilon < 0$.
Thus, there emerges the Type II (i.e., sudden) singularity.
Additionally, when $\mathcal{G} \to -0$,
$F(\mathcal{G}) \sim l_4 |\mathcal{G}|^{v}$,
where $l_4 (<0)$ a negative constant and $1<v<\infty$,
we get $-1 < \upsilon < -1/3$.
As a result, the Type II singularity appears.
If for $\mathcal{G} \to -0$,
$F(\mathcal{G}) \sim l_3 |\mathcal{G}|^{v}$, where $2/3 < v < 1$
and $v \neq 2m /\left(3m - 1\right)$ with
$m = 1, 2, 3, \dots$ a natural number,
we see that $-\infty < \upsilon < -1$.
Consequently, there occurs the Type IV singularity.
We also remark that for
$H = \bar{h}/\left( t_{\mathrm{s}} -t \right)^{-1/3}$,
there can appear any kind of the Type II singularity.
For it, we have
$\mathcal{G} = 24 \bar{h}^{3} \upsilon + 24 \bar{h}^{4}\left( t_{\mathrm{s}} -t \right)^{4/3} < 0$.
Eventually,
for $F(\mathcal{G})= j_1 \mathcal{G} \sqrt{\mathcal{G}+ r_1} + j_2
\sqrt{\mathcal{G} + r_1}$ with
where $j_1 (>0)$, $j_2 (>0)$ and $r_1 (>0)$
are positive constants, the Type II singularity happens.

\textit{Examples of $F(\mathcal{G})$ gravity realizing bounce cosmology}--
%
We present several simple examples of bounce cosmology.

\noindent
{\bf Case (i): Exponential model}

We examine the following form of the scale factor
\begin{equation}
a(t)=\exp\left(\tilde{\alpha} t^2\right)\,.
\label{eq:47}
\end{equation}
In this case, the functions $P(t)$ and $Q(t)$ are represented as
\begin{eqnarray}
P(t) \Eqn{=} s_1+s_2 \left(-\frac{\e^{t^2 \tilde{\alpha} }}{t}+ \sqrt{\pi }\sqrt{-\tilde{\alpha} }\text{Erf}\left(t \sqrt{-\tilde{\alpha} }\right)\right)+\frac{ \sqrt{\pi } \sqrt{-\tilde{\alpha} } \text{MeijerG}\left[\{\{0\},\{1\}\},\left\{\{0,0\},\left\{-\frac{1}{2}\right\}\right\},-t^2 \tilde{\alpha} \right]}{16 (-\tilde{\alpha}) ^{3/2}}\,,
\label{eq:48} \\
Q(t) \Eqn{=} -24 t \tilde{\alpha} \left[\tilde{\alpha} \left(t+8 s_2 \e^{t^2 \tilde{\alpha} } \tilde{\alpha} \right)+\e^{t^2 \tilde{\alpha} } \sqrt{\pi } \sqrt{-\tilde{\alpha} } \text{Erf}\left(t \sqrt{-\tilde{\alpha} }\right)
\right]\,,
\label{eq:49}
\end{eqnarray}
with $s_1$ and $s_2$ constants.
Here, $\tilde{\alpha} (<0)$ is a negative constant, ${\rm Erf}(x)=\left(2/\sqrt{\pi}\right) \int_0^x \e^{-t^2}dt$, and
$\text{MeijerG}[\{a_1,..,a_p\},\{b_1,..,b_q\},z]$ is the Meijer G function
$G_{pq}^{mn}\left(z\left|
\begin{array}{c}
  a_1,\ldots ,a_p \\
  b_1,\ldots ,b_q
\end{array}
\right.\right)$.
Thus, we find
\begin{eqnarray}
F(\mathcal{G}) \Eqn{=}
s_1 \mathcal{G}-s_2\left(\frac{48}{\sqrt{3}}\tilde{\alpha}^2
\mathcal{X}_1 \mathcal{X}_2
+4\sqrt{3}\mathcal{G}\tilde{\alpha} \frac{\mathcal{X}_1}{\mathcal{X}_2}
+\sqrt{\pi}\mathcal{G}\sqrt{-\tilde{\alpha}} \mathcal{X}_3 \right)
+\frac{1}{2} \left( - \mathcal{X}_2^2 \right)
-\sqrt{2}\sqrt{-\tilde{\alpha}}
\mathcal{X}_2 \mathcal{X}_1 \sqrt{\pi} \mathcal{X}_3
\nonumber \\
&&
{}+\frac{\sqrt{\pi}\mathcal{G}}{16 \tilde{\alpha}}\text{MeijerG}\left[\{\{0\},\{1\}\},\left\{\{0,0\},\left\{-\frac{1}{2}\right\}\right\},\frac{1}{4}-\frac{\sqrt{\mathcal{G}+24\tilde{\alpha}^2}}{8 \sqrt{6}\tilde{\alpha}}\right]\,.
\label{eq:50}
\end{eqnarray}
with
\begin{eqnarray}
\mathcal{X}_1 \Eqn{\equiv}
\exp \left( -\frac{1}{4}+\frac{\sqrt{\mathcal{G}+24\tilde{\alpha}^2}}{8\sqrt{6}\tilde{\alpha}} \right)\,,
\label{eq:51} \\
\mathcal{X}_2 \Eqn{\equiv}
\sqrt{-12\tilde{\alpha} +\sqrt{6}\sqrt{\mathcal{G}+24 \tilde{\alpha}^2}}\,,
\label{eq:52} \\
\mathcal{X}_3 \Eqn{\equiv}
\text{Erf}\left(\frac{\sqrt{-12\tilde{\alpha}+\sqrt{6}\sqrt{\mathcal{G}+24\tilde{\alpha}^2}}}{4\sqrt{3}\sqrt{-\tilde{\alpha}}}\right)\,.
\label{eq:53}
\end{eqnarray}
The form in Eq.~(\ref{eq:50}) is defined for $\tilde{\alpha} <0$ and $\mathcal{G} \geq -24 \tilde{\alpha}^2$.
It is evident that
with this choice of the scale factor,
there does not arise any kind of singularities,
that is, the scale factor, the Hubble parameter, and its derivative
are finite.
A similar situation takes place for the metric in Eq.~(\ref{eq:6}),
because the parameters $\tau$ and $\sigma$ have the same sign
with each other.
For the reconstruction of $F(R)$ gravity,
there exists a solution with different signs of $\tau$ and $\sigma$.
In this case, there exists a situation in which the scale factor vanishes,
and the Hubble parameter and its derivative tends to infinity
(the type III singularities).
However, we can construct an example of cosmology with singularities
in $F(\mathcal{G})$ gravity.

\noindent
{\bf Case (ii): Power-law model}

We investigate the scale factor with its form
\begin{equation}
a(t)= \beta t^{2n} \,,
\label{eq:54}
\end{equation}
where $\beta$ is a constant and $n$ is an integer.
In this case, the functions $P(t)$ and $Q(t)$ become
\begin{eqnarray}
P(t) \Eqn{=} y_1+y_2\frac{t^{3+2n}}{3+2n}-\frac{t^2}{8n\left(1+2n\right)} \,,
\label{eq:55} \\
Q(t) \Eqn{=}
\frac{24 n^2 \left[1-2 n-8 y_2 n \left(1+2 n \right) t^{1+2 n}\right]}{\left(1+2 n \right) t^2}\,,
\label{eq:56}
\end{eqnarray}
where $y_1$ and $y_2$ are integration constants.

The cosmic time is expressed in terms of the Gauss-Bonnet invariant as
\begin{equation}
t=\pm 2 \sqrt{2}\,\, 3^{1/4} \left[\frac{n^3 (-1+2 n)}{\mathcal{G}}
\right]^{1/4}\,.
\label{eq:57}
\end{equation}
Thus, the final form of $F(\mathcal{G})$ can be written as
\begin{equation}
F(\mathcal{G}) = y_1\mathcal{G}+y_2\mathcal{G}^{\left(1-2n\right)/4}
-\frac{2\sqrt{3}\sqrt{n^3 \left(-1+2n \right)
\mathcal{G}}}{n \left(1+2n\right)}\,.
\label{eq:58}
\end{equation}
It is clearly seen that the parameter $n$ can be both positive and negative values. For example, if $n=1$, we have
\begin{equation}
F_1(\mathcal{G})=-\frac{y_2}{\mathcal{G}^{1/4}}-\frac{2 \sqrt{\mathcal{G}}}{\sqrt{3}}+y_1 \mathcal{G} \,,
\label{eq:59}
\end{equation}
whereas for $n=-1$, we find
\begin{equation}
F_2(\mathcal{G})=-6 \sqrt{\mathcal{G}}-y_2 \mathcal{G}^{3/4}
+y_1 \mathcal{G}\,.
\label{eq:60}
\end{equation}
In this model, we acquire the type III singularity.
We mention that it is possible to examine more general case that
the scale factor is described as
$a(t) = z_1 t^{2n}+z_2$ with $z_1$ and $z_2$ constants,
but that for this model, $F(\mathcal{G})$ would become quite complicated.

\noindent
{\bf Case (iii): Ekpyrotic scenario}

The so-called ekpyrotic universe is an alternative explanation to the inflationary paradigm proposed one decade ago in Ref.~\cite{Enc}. 
It can provide a realistic picture of the universe evolution (for a confrontation between both models, see~\cite{Enc2}). 
In the same way as the inflationary scenario, the ekpyrotic cosmological models can also predict the origin of primordial inhomogeneities that leads to the formation of large structures an d the anisotropies observed in the CMB radiation. In addition, this model does not require initial conditions in comparison with the standard inflationary scenario due to 
its cyclic nature (see, for example,~\cite{Enc1}). 

Let us now consider a model that may reproduce a entire cycle of an ekpyrotic universe
\begin{equation}
H(t)=H_0-H_1 \e^{-\beta t},\,\,\, H_0>0,\,\,\, \beta >0 \,, 
\label{eq:Ekp-}
\end{equation}
where $H_0$, $H_1$, and $\beta$ are constants. 
In this case, the scale factor takes the form 
\begin{equation}
a(t)= a_0 \e^{\e^{-\beta t}\frac{ H_1}{\beta}+H_0 t}  \,,
\label{eq:Ekp-2}
\end{equation}
where $a_0$ is a constant of integration. 
Moreover, the functions $P(t)$ and $Q(T)$ can only be found 
in the form of integrals 
\begin{eqnarray}
P(t) \Eqn{=} c_1+\int^t_1 dz_2\left(
\frac{\e^{\frac{H_1 }{\beta } \e^{-\beta  z_2}+H_0 z_2} c_2}{\left(H_0-\e^{-\beta  z_2} H_1\right)^2}\left(1-
\int^{z_2}_1 dz_1 
\e^{-\frac{e^{-\beta   z_1} H_1}{\beta }-H_0  z_1+\beta   z_1}\frac{ H_1 \left(H_0-\e^{-\beta   z_1} H_1\right)^2 \beta }{2 \left(\e^{\beta  z_1}H_0-H_1\right)^2} 
\right)
\right)\,,
\label{eq:Ekp-3} \\
Q(t) \Eqn{=} -6 \e^{-2 t \beta } \left(\e^{t \beta } H_0-H_1\right) \left(2 \e^{H_0 t+\frac{\e^{-t \beta } H_1}{\beta }+t \beta }
\left(2c_2-\int^t_1 dz_1
 \e^{-\frac{\e^{-z_1 \beta } H_1}{\beta }-z_1 (H_0+\beta )}H_1 \beta
\right)+\e^{t \beta }  H_0- H_1\right)\,.
\label{eq:Ekp-4}
\end{eqnarray}
These expressions have been obtained under conditions $H_0>0$ and $\beta>0$, 
and $c_1$ and $c_2$ are constants of integration. 

We now find the time dependence of $G$ as 
\begin{equation}
t=\frac{\text{ln}\left[\frac{12 H_1^2}{\sqrt{6} \sqrt{\mathcal{G}}+12 H_0^2}\right]}{8 H_0}\,.
\label{eq:Ekp-5}
\end{equation}
This expression is defined for $G>24 H_0^4$. 
\begin{eqnarray}
F(\mathcal{G}) \Eqn{=} c_1 \mathcal{G}+c_2 \left(24 \e^{\frac{\e^{-4 H_0 t} H_1}{4 H_0}+H_0 t}\left(\e^{-4 H_0 t} H_1-H_0\right)+\mathcal{G}\int _1^t\frac{\e^{\frac{\e^{-4 H_0 z_2} H_1}{4 H_0}+9 H_0 z_2}}{\left(-\e^{4 H_0 z_2} H_0+H_1\right)^2}dz_2\right) \nonumber\\
&&{}-6\left(\e^{-4 H_0 t} H_1-H_0\right)^2-48 H_0 H_1 \e^{\frac{\e^{-4 H_0 t} H_1}{4 H_0}+ H_0 t} \left(\e^{-4 H_0 t} H_1-H_0\right)\int _1^t \e^{-\frac{\e^{-4 H_0 z_1} H_1}{4 H_0}-5 H_0 z_1} dz_1 \nonumber\\
&&{}-\mathcal{G}\int_1^t \left(\int _1^{z_2}\frac{2 H_0 H_1}{\left(\e^{4 H_0 z_2} H_0-H_1\right)^2}\e^{\frac{\e^{-4 H_0 z_2} H_1}{4 H_0}+9 H_0 z_2}\e^{-\frac{\e^{-4 H_0 z_1} H_1}{4 H_0}-5 H_0 z_1}dz_1\right) \, dz_2 \,.
\end{eqnarray}
Thus, the ekpyrotic cosmology may be realized naturally in $F(\mathcal{G})$ gravity.

\textit{Sum of multiple exponential functions model}--
%
We explore a sum of multiple exponential functions model of the scale factor
\begin{equation}
a(t) = \exp \left( \alpha t^2 \right) + \exp \left( \alpha^2 t^4 \right)\,,
\quad
\alpha >0\,,
\label{eq:31}
\end{equation}
with $\alpha$ a positive constant. {}From this expression, we have
\begin{equation}
H(t) = \frac{2\alpha t \left(1+2\alpha t^2 \exp \left( \alpha^2 t^4 -\alpha t^2 \right) \right)}{1+ \exp \left( \alpha^2 t^4 -\alpha t^2 \right)}\,.
\label{eq:32}
\end{equation}
Here, we note that $\mathcal{G} (t) =24H^2 \left(\dot{H}+H^2 \right) \geq 0$
for $t \in \mathbb{R}$.
We have the following equation
\begin{equation}
p_2 (\mathcal{G}) \frac{d^2 F(\mathcal{G})}{d \mathcal{G}^2}
+p_1 (\mathcal{G}) \frac{d F(\mathcal{G})}{d \mathcal{G}}
+F(\mathcal{G}) = b(\mathcal{G})\,,
\label{eq:33}
\end{equation}
%
where the coefficients are presented in the form of series in powers of $\mathcal{G}$ in the neighborhood of $\mathcal{G} = 0$
\begin{eqnarray}
p_2 (\mathcal{G}) \Eqn{=}
2\mathcal{G}^2 + \frac{5}{8 \alpha^2}\mathcal{G}^3 -\frac{113}{144 \alpha^4} \mathcal{G}^4 + o(\mathcal{G}^4)\,,
\label{eq:34} \\
p_1 (\mathcal{G}) \Eqn{=} -\mathcal{G}\,,
\label{eq:35} \\
b (\mathcal{G}) \Eqn{=} -\frac{1}{4\alpha}\mathcal{G} + \frac{17}{192 \alpha^3}\mathcal{G}^2 -\frac{379}{4608 \alpha^5} \mathcal{G}^3 + o(\mathcal{G}^3)\,.
\label{eq:36}
\end{eqnarray}
We seek a solution of equation (\ref{eq:33}) in the neighborhood of $\mathcal{G} = 0$.
First of all, we construct a fundamental system of solutions of the corresponding homogeneous equation
\begin{equation}
\frac{d^2 F(\mathcal{G})}{d \mathcal{G}^2}
+q_1 (\mathcal{G}) \frac{d F(\mathcal{G})}{d \mathcal{G}}
+q_2 (\mathcal{G}) F(\mathcal{G}) = 0\,.
\label{eq:37}
\end{equation}
%


Since the coefficients $q_k (\mathcal{G})$ for $k=1, 2$ has a pole of order not higher than $k$ at $ G = 0$, then $G = 0$ is regular singular point of equation (\ref{eq:37}). Therefore, we can obtain
\begin{equation}
q_1 (\mathcal{G}) = \frac{\bar{q}_1 (\mathcal{G})}{\mathcal{G}}\,,
\quad
q_2 (\mathcal{G}) = \frac{\bar{q}_2 (\mathcal{G})}{\mathcal{G}^2}\,.
\label{eq:38}
\end{equation}
%

Here, $\bar{q}_1 (\mathcal{G})$ and $\bar{q}_2 (\mathcal{G})$ are
holomorphic functions in a neighborhood of $\mathcal{G} = 0$ .
We construct a fundamental system of solutions of Eq.~(\ref{eq:37})
in the neighborhood of $\mathcal{G} = 0$.
Solutions will be found in the form of a generalized series
\begin{equation}
F_1 (\mathcal{G}) = \mathcal{G}^\gamma \sum_{k=0}^{+ \infty} c_k \mathcal{G}^k
\,.
\label{eq:39}
\end{equation}
By combining this expression with Eq.~(\ref{eq:37}), we acquire
\begin{equation}
\gamma \left( \gamma -1 \right) + \bar{q}_1 (0) \gamma + \bar{q}_2 (0) = 0\,.
\label{eq:40}
\end{equation}

We find the following solutions of this equation:
$\gamma_1 = 1/2$ and $\gamma_2 = 1$. {}From Eq.~(\ref{eq:37}), we represent
\begin{equation}
F_1 (\mathcal{G}) = \mathcal{G} \varphi_1 (\mathcal{G})\,,
\quad
F_2 (\mathcal{G}) = \sqrt{\mathcal{G}} \varphi_2 (\mathcal{G})\,,
\label{eq:41}
\end{equation}
%
where $\varphi_k (\mathcal{G})$ for $k=1, 2$ are holomorphic functions in $\mathcal{G}=0$ at that $\varphi_k (\mathcal{G}) \neq 0$.
Substituting (\ref{eq:40}) into the homogeneous equation,
we obtain a recurrent system from which we consistently find the coefficients
$c_1$, $c_2$, $\cdots$
Thus, the fundamental system of the homogeneous equation has the form
\begin{eqnarray}
F_1 (\mathcal{G}) \Eqn{=} \mathcal{G}\,,
\label{eq:42} \\
F_2 (\mathcal{G}) \Eqn{=} \sqrt{\mathcal{G}} \left(1+\frac{5}{32\alpha^2}\mathcal{G} -\frac{2483}{55296\alpha^4}\mathcal{G}^2 \right) + o(\mathcal{G}^3)\,.
\label{eq:43}
\end{eqnarray}
%
Solving the inhomogeneous equation (\ref{eq:33}) by the method of variation
of constants,
we obtain the particular solution of the inhomogeneous equation
\begin{equation}
F_3 (\mathcal{G}) =
\frac{1}{2\alpha} \left(1-\frac{1}{2} \ln \mathcal{G} \right)\mathcal{G}
+\frac{47}{576\alpha^3}\mathcal{G}^2
-\frac{1753}{46080\alpha^5}\mathcal{G}^3
+ o(\mathcal{G}^3)\,.
\label{eq:44}
\end{equation}
%
As a result, an approximate solution of Eq.~(\ref{eq:33}) has the form
\begin{equation}
F(\mathcal{G}) = c_1 \mathcal{G} + c_2 \sqrt{\mathcal{G}}
\left( 1+\frac{5}{32\alpha^2}\mathcal{G}
-\frac{2483}{55296\alpha^4}\mathcal{G}^2 \right)
+\frac{1}{2\alpha} \left(1-\frac{1}{2} \ln \mathcal{G} \right)\mathcal{G}
+\frac{47}{576\alpha^3}\mathcal{G}^2
+ o(\mathcal{G}^3)\,.
\label{eq:45}
\end{equation}
%

\textit{Unification of bounce with the late-time cosmic acceleration}--
%
We study the reconstruction of an $F(\mathcal{G})$ gravity theory where both
the bouncing behavior in the early universe and the
late-time accelerated expansion of the universe at the dark energy
dominated stage can occur within a unified model.

\begin{center}
\begin{figure}[tbp]
\resizebox{!}{6.5cm}{
    \includegraphics{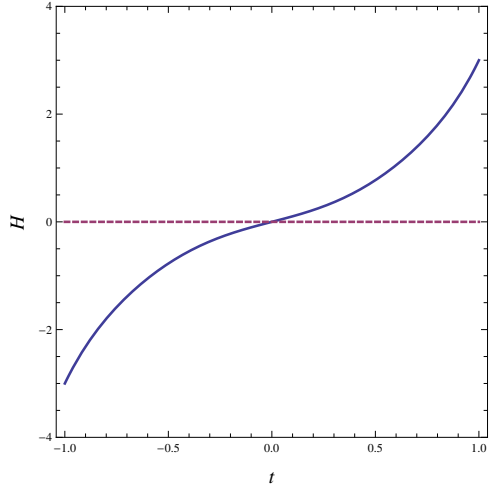}
                   }
\caption{The behavior of the Hubble parameter in Eq.~(\ref{eq:32})
with $\alpha = 1/t_\star^2 =1$ (where we take $t_\star =1$)
around $t=0$. The dotted line draws $H=0$.
}
\label{Fig-1}
\end{figure}
\end{center}

In Eq.~(\ref{eq:31}), we take $\alpha = 1/t_\star^2$ with
$t_\star$ a fiducial time~\cite{Bamba:2013fha}. We also have
$N = \ln a/a_\star$ with $a_\star = 1$
and hence $H = \dot{N}$. {}From Eq.~(\ref{eq:32}),
we see that for $t<0$, $H <0$, whereas for $t>0$, $H>0$,
so that around $t=0$, namely, in the early universe,
the bouncing behavior of the universe can be realized.
In Fig.~\ref{Fig-1}, we show the behavior of the Hubble parameter
in Eq.~(\ref{eq:32}) with $\alpha = 1/t_\star^2 =1$,
where $t_\star =1$, around the bouncing time $t=0$.
Clearly, it can be observed that the value of $H$ evolves from
negative to positive as the cosmic time $t$ does. 

On the other hand, when $\alpha t^2 \gg 1$, the universe
can be considered to be at the dark energy dominated stage,
because for $\alpha t^2 \gg 1$, we find
$a(t) \approx \exp \left( \alpha^2 t^4 \right)$.
Eventually, we obtain
\begin{equation}
\ddot{a} (t) \approx 4 \alpha^2 t^2 \left(3+4\alpha^2 t^4 \right)
\exp \left( \alpha^2 t^4 \right) >0\,.
\label{eq:46}
\end{equation}
This implies the accelerated expansion of the universe happens.
As a result, we see that in the case that
the scale factor is given by Eq.~(\ref{eq:31}),
which is expressed as a sum of two exponential functions,
the late-time cosmic acceleration as well as the bouncing
behavior in the early universe can be realized in a unified
manner. For $a(t)$ in Eq.~(\ref{eq:31}), the form of
$F(\mathcal{G})$ is represented as in Eq.~(\ref{eq:45}).

In addition, if $a(t)$ is given by Eq.~(\ref{eq:31}),
the stability conditions
$\mathcal{J}_2/\mathcal{J}_1 >0$ and $\mathcal{J}_3/\mathcal{J}_1 >0$
with Eqs.~(\ref{eq:22})--(\ref{eq:24}) can be satisfied.
For $\alpha t^2 \ll 1$, namely, around the bounce in the early universe, 
we have $a(t) \approx \exp \left( \alpha t^2 \right)$, 
$H \approx 2 \alpha t$, $N \approx \alpha t^2 \ll 1$, 
$\tilde{g}_0 (N) \approx 4N$, and
$\mathcal{G}_0 \approx 192N\left(1+2N\right) \approx 192N$. 
Hence, for $F(\mathcal{G}_0)$ in Eq.~(\ref{eq:45}), we find 
\begin{eqnarray}
\frac{\mathcal{J}_2}{\mathcal{J}_1} \Eqn{\approx} \frac{3}{2} 
\left[ 2+\frac{1}{N} + 128\left(1+4N\right) \frac{F'''(\mathcal{G}_0)}{F''(\mathcal{G}_0)} \right] \approx 3 >0\,,
\label{eq:61} \\
\frac{\mathcal{J}_3}{\mathcal{J}_1} \Eqn{\approx} 
\frac{1}{3072N^3} 
\left\{ \frac{1}{F''(\mathcal{G}_0)} +1536N \left[ \left(3+18N-8N^2 \right) 
+384N\left(1+4N\right)^2 \frac{F'''(\mathcal{G}_0)}{F''(\mathcal{G}_0)} 
\right] 
\right\}\,. 
\label{eq:62}
\end{eqnarray}
Regarding $\mathcal{J}_3/\mathcal{J}_1$ in Eq.~(\ref{eq:62}), 
only for the value of the content within the brackets $[\,\,\,]$ is 
positive, we have $\mathcal{J}_3/\mathcal{J}_1 >0$, so that 
the bouncing solution can be stable. 
While for $\alpha t^2 \gg 1$, namely, in the late-time universe, 
we have $a(t) \approx \exp \left( \alpha^2 t^4 \right)$, 
$H \approx 4 \alpha^2 t^3$, $N \approx \alpha^2 t^4 \gg 1$, 
$\tilde{g}_0 (N) \approx 16\alpha N^{3/2}$, and
$\mathcal{G}_0 \approx 1536 \alpha^2 N^2\left(4N+3\right)$. 
Thus, for $F(\mathcal{G}_0)$ in Eq.~(\ref{eq:45}), we obtain 
\begin{eqnarray}
\frac{\mathcal{J}_2}{\mathcal{J}_1} \Eqn{\approx} 
3 N^{1/2} \left[ 
N +\frac{3}{4} + 3072 \alpha^3 N^2 \left( 2N+1 \right) \frac{F'''(\mathcal{G}_0)}{F''(\mathcal{G}_0)} \right] 
\approx 3 N^{1/2} \left( N+\frac{3}{4} + \frac{2N+1}{4N+3} \right) >0\,, 
\label{eq:63} \\
\frac{\mathcal{J}_3}{\mathcal{J}_1} \Eqn{\approx} 
\frac{1}{196608 \alpha^3 N^3} 
\left[ \frac{1}{F''(\mathcal{G}_0)} + 24576 \alpha^3 N^{5/2} \left(-32N^2+108N+45 \right)  
\right. 
\nonumber \\ 
&&
{} +\left. 
905969664 \alpha^5 N^{9/2} 
\left(8N+3\right)\left(2N+1\right) 
\frac{F'''(\mathcal{G}_0)}{F''(\mathcal{G}_0)} \right]
\nonumber \\
\Eqn{\approx} 
\frac{-32N^2+108N+45}{8N^{1/2}}\,.
\label{eq:64}
\end{eqnarray}
Here, the last approximate equality in Eq.~(\ref{eq:64}) follows from 
the approximate relations 
$1/F''(\mathcal{G}_0) \approx -221184 \alpha^4/\left[ 2483\left( 8-c_2 \right)\right] \mathcal{G}_0^{-1/2}$
and 
$F'''(\mathcal{G}_0)/F''(\mathcal{G}_0) \approx \left(3c_2 +1 \right)/\left[ 2\left( 8-c_2 \right)\right] \mathcal{G}_0^{-1}$. 
That is, since $\mathcal{G}_0 \gg 1$, the second term in the brackets 
$[\,\,\,]$ on the right-hand side of the first approximate equality 
in Eq.~(\ref{eq:64}) would be the dominant term. 
If $0< N <15/4$, we see that $\mathcal{J}_3/\mathcal{J}_1 >0$ , and 
therefore the solution in the late-time universe can be stable. 
We mention that among three terms in the brackets $[\,\,\,]$, 
when the two or three terms would be comparable with each other, 
by taking into consideration the contributions from 
not only the second term but also the other ones, 
even for larger $N$, the condition $\mathcal{J}_3/\mathcal{J}_1 >0$ 
could be met.

\textit{Summary}--
%
We have studied bounce cosmology in $F(\mathcal{G})$ gravity.
We have reconstructed $F(\mathcal{G})$ gravity model with
the bouncing behavior in the early universe.
Also, we have analyzed the stability of the solutions in
the reconstructed model.
Moreover, we have explored an exponential model and a power-law model
and found that in these models the bouncing behavior can happen. 
Furthermore, the $F(\mathcal{G})$ gravity theory with 
the ekpyrotic scenario has been investigated. 
In addition, it has been verified that in a sum of two exponential functions model of the scale factor, the unification of the bounce in the early universe
with the current cosmic acceleration can be realized. 
In this unified model, we have further examined the stability of 
the bounce and late-time solutions. 

We here mention the comparison of this unified scenario with the observations by following the discussions in Ref.~\cite{Bamba:2012cp}. 
It is known that the cosmography can be adopted to test modified gravity theories such as $F(R)$ gravity. Therefore, by analogy with this fact, 
it is considered that with the cosmographical procedure 
in which the Hubble, deceleration, jerk, snap, and lerk parameters are 
examined, our unified model of the early-time bounce 
with the late-time cosmic acceleration in $F(\mathcal{G})$ gravity 
can also be checked whether it can be consistent with the recent observational data. This point is regarded as the strong advantage of the cosmography. 

The novel significant ingredient observed in this work is that
we have explicitly derived the $F(\mathcal{G})$ gravity model analytically,
where the early-time bouncing behavior and the late-time cosmic acceleration
can occur within the framework of the single model.
Such behaviour supports the first proposal on the unification of 
early-time and 
late-time accelerations within modified gravity as given in Ref.~\cite{NO}. 
This analysis is considered to be a useful clue for building models to
represent the early universe, which should be described by the high-energy
theories, and for seeking for the cosmological mechanism of the current accelerated expansion of the universe.

\textit{Acknowledgments}--
K.B. would like to sincerely appreciate very kind support of
Prof. Gi-Chol Cho, Prof. Shin'ichi Nojiri, Prof. Akio Sugamoto
and Prof. Koichi Yamawaki.
The work is supported in part
by the JSPS Grant-in-Aid for
Young Scientists (B) \# 25800136 (K.B.),
and
the grant of Russian Ministry of Education and Science, project TSPU-139 and
the grant for LRSS, project No 88.2012.2 (S.D.O. and A.N.M.).


\end{document}